\documentclass[12pt]{article}
\pdfoutput=1
\usepackage{putex}
\usepackage{feyn}
\usepackage[vcentermath]{youngtab}
\usepackage{subfig}
\usepackage{lscape}

\usepackage{graphicx}
\usepackage{epstopdf}
\usepackage{enumerate}
\usepackage{cite}
\usepackage{tensor}
\usepackage{slashed}
\usepackage{amsmath}
\usepackage{amssymb}
\usepackage{mathrsfs}
\usepackage{lgrind}

\usepackage{hyperref}

\numberwithin{equation}{section}

\newcommand {\be} {\begin {equation}}
\newcommand {\ee} {\end {equation}}

\newcommand {\bes} {\begin {equation*}}
\newcommand {\ees} {\end {equation*}}

\newcommand{\eps}{\epsilon}

\def\CH{{\cal H}}

\newcommand{\beq}{\begin{equation}}
\newcommand{\eeq}{\end{equation}}

\def\be{ \begin{equation} }
\def\ee{ \end{equation} }

\begin{document}

\institution{IAS}{Institute for Advanced Study, Princeton, NJ 08540, USA}

\title{
Critical Field Theories with OSp$(1|2M)$ Symmetry 
}

\authors{Igor R.~Klebanov\worksat{\IAS}\footnote{On leave from Princeton University.} 
}

\abstract{
In the paper [L. Fei et al., JHEP {\bf 09} (2015) 076] a cubic field theory of a scalar field $\sigma$ and two anticommuting scalar fields, $\theta$ and $\bar \theta$, was formulated. In $6-\epsilon$ dimensions it has a weakly coupled fixed point with imaginary cubic couplings where the symmetry is enhanced to the supergroup OSp$(1|2)$. This theory may be viewed as a ``UV completion" in $2<d<6$ of the non-linear sigma model with hyperbolic target space H$^{0|2}$ described by a pair of intrinsic anticommuting coordinates. It also describes the $q\rightarrow 0$ limit of the critical $q$-state Potts model, which is equivalent
to the statistical mechanics of spanning forests on a graph. 
In this letter we generalize these results to a class of OSp$(1|2M)$ symmetric field theories whose upper critical dimensions are $d_c(M) = 2 \frac{2M+1}{2M-1}$. 
They contain $2M$ anticommuting scalar fields, $\theta^i, \bar \theta^i$, and one commuting one, with interaction $g\left (\sigma^2+ 2\theta^i \bar \theta^i \right )^{(2M+1)/2}$.
In $d_c(M)-\epsilon$ dimensions, we find a weakly coupled IR fixed point at an imaginary value of $g$. We propose that these critical theories are the UV completions of the sigma models with fermionic hyperbolic target spaces H$^{0|2M}$.
Of particular interest is the quintic field theory with OSp$(1|4)$ symmetry, whose upper critical dimension is $10/3$. Using this theory, we make a prediction for the critical behavior of the 
OSp$(1|4)$ lattice system
in three dimensions.
 }

\date{}
\maketitle

%\tableofcontents

\section{Introduction}

This letter builds on the paper \cite{Fei:2015kta} where the field theory was studied 
with the Euclidean action
\be
\label{cubicOSp}
S=\int d^d x \bigg (
\partial_\mu \theta \partial^\mu \bar \theta + \frac{1}{2}\left(\partial_{\mu}\sigma\right)^2 + g_1 \sigma \theta \bar \theta + \frac{1}{6}g_2 \sigma^3
\bigg )
\ee
for two anticommuting scalar fields, $\theta$ and $\bar \theta$, and one commuting one, $\sigma$.
The global Sp$(2)$ symmetry of this model
becomes enhanced to the supergroup
OSp$(1|2)$ because at the IR fixed point in $6-\epsilon$ dimensions the two coupling constants are {\it imaginary} and related by $g_2^*= 2 g_1^*$. At this weakly coupled fixed point,
which is of the Wilson-Fisher type \cite{Wilson:1971dc},
the interaction becomes proportional to the manifestly OSp$(1|2)$ invariant form $(\sigma^2+ 2\theta \bar \theta)^{3/2}$.
Using 3-loop calculations, the scaling dimensions of $\sigma$ and $\theta$ in $6-\eps$ dimensions were indeed found to be equal \cite{Fei:2015kta}. Since the coupling is imaginary, the Euclidean path integral associated with (\ref{cubicOSp}) does not suffer from the instability encountered for the real $\sigma^3$ interaction. Indeed, the cubic theory of a single scalar field with an imaginary coupling constant is known \cite{Fisher:1978pf} to describe the Lee-Yang edge singularity. 

Earlier examples of models with OSp$(1|2)$ symmetry are the lattice models and sigma models describing the spanning forests, or equivalently the $q\rightarrow 0$ limit of the $q$-state Potts model
\cite{Caracciolo:2004hz,Jacobsen:2005uw,2007JPhA...4013799C}.
In \cite{Fei:2015kta} we showed that the $6-\eps$ expansions of the scaling dimensions in our OSp$(1|2)$ symmetric theory are the same as in the 
$q\rightarrow 0$ limit of the $q$-state Potts model \cite{deAlcantaraBonfim:1981sy}. Thus, (\ref{cubicOSp}) provides an explicit formulation of the field theory that governs this formal limit and is super-renormalizable in $d<6$.

Besides the $6-\eps$ expansion, it is interesting to develop the $2+\eps$ expansion for the critical theory of spanning forests. In \cite{Caracciolo:2004hz,Caracciolo:2017xyc} it was argued that it is provided by
the OSp$(1|2)$ sigma model \cite{Read:2001pz,Saleur:2003zm} with the action
\be
\label{sigmaOSp}
S=\frac {1} {2 g^2}\int d^d x \bigg ( (\partial_\mu \sigma)^2 +
2\partial_\mu \theta \partial^\mu \bar \theta 
%+ \lambda (\sigma^2 + \theta \bar \theta -1)
\bigg )
\ ,\ee
where the constraint $\sigma^2 + 2\theta \bar \theta =1$ is imposed. The constraint has two solutions, $\sigma=\pm (1-\theta \bar \theta)$, and
choosing one of these solutions breaks the $Z_2$ symmetry but preserves the OSp$(1|2)$; then the global
symmetries in the sigma model are the same as in the cubic theory (\ref{cubicOSp}). 
In the recent work \cite{2021CMaPh.381.1223B,2021arXiv210701878B} the interpretation of the target space was changed from (half of) the sphere S$^{0|2}$ to the space H$^{0|2}$,
which is a fermionic version of the hyperboloid. 
%because $g^2$ has to be made negative.

Substituting $\sigma=1-\theta \bar \theta$ into (\ref{sigmaOSp}), we find the sigma model with the hyperbolic target space H$^{0|2}$. It has the following classical action in terms of the two anticommuting scalar fields:
\be
\label{nonlinear}
S=\frac {1} {g^2}\int d^d x \bigg (
\partial_\mu \theta \partial^\mu \bar \theta -  \theta \bar \theta \partial_\mu \theta \partial^\mu \bar \theta\bigg )
\ ,\ee
and it is important to take $g^2<0$ so that the model is asymptotically free in $d=2$ \cite{Caracciolo:2004hz,2021CMaPh.381.1223B,2021arXiv210701878B}.
\footnote{In theories with anticommuting fields, rigorous approaches to Renormalization Group are sometimes available \cite{Giuliani:2020aot}.} 
 Indeed, 
the beta function of this theory is the same as for the O$(N)$ non-linear sigma model continued to $N=-1$, and the theory is asymptotically free for negative $g^2$. Thus, in $d=2+\eps$ this theory has a weakly coupled UV fixed point.
The fact that near the upper critical dimension $6$ the model has another weakly coupled description, involving an extra canonical field $\sigma$, is analogous to a similar phenomenon in the Gross-Neveu model
\cite{Gross:1974jv}
near $4$ dimensions \cite{ZinnJustin:1991yn, Hasenfratz:1991it} (for a more recent discussion, see \cite{Fei:2016sgs}).

We may interpret the cubic theory (\ref{cubicOSp}) as a ``UV completion" of the H$^{0|2}$ sigma model in $2<d<6$: the IR fixed point of the cubic theory presumably describes the same physics as the UV fixed point of the sigma model. The sense of this is similar to how the super-renormalizable Gross-Neveu-Yukawa model \cite{ZinnJustin:1991yn, Hasenfratz:1991it} provides a UV completion in $2<d<4$ of the non-renormalizable  Gross-Neveu model. Similarly, the $O(N)$ symmetric field theory of $N>2$ scalar fields with the quartic interaction 
$\lambda (\phi^i \phi^i)^2$  provides a UV completion in $2<d<4$ of the non-renormalizable $O(N)$ non-linear sigma model.

The OSp$(1|2)$ lattice system \cite{Caracciolo:2004hz,Jacobsen:2005uw,2007JPhA...4013799C,2021CMaPh.381.1223B,2021arXiv210701878B}, which underlies the continuum descriptions reviewed above, involves introducing on each lattice site $x$ the vector $u_x=(\theta_x, \bar \theta_x,\sigma_x)$. The constraint that it belong to
H$^{0|2}$ implies $\sigma_x=1- \theta_x \bar \theta_x$. Then for each pair of nearest neighbor lattice sites, $x$ and $y$, the factor
\be
e^{\beta (u_x\cdot u_y+1)/2}\ , \qquad u_x\cdot u_y = - \sigma_x \sigma_y - \theta_x  \bar \theta_y -  \theta_y  \bar \theta_x \ ,
\ee
is included in the integrand of the partition function. Since all the integrations are over the Grassmann variables $\theta_x, \bar \theta_x$, they can be performed exactly on a finite lattice. This lattice system has a particular simplicity and 
is found to be equivalent to the statistical mechanics of spanning forests, or alternatively the $q\rightarrow 0$ Potts model \cite{Caracciolo:2004hz,Jacobsen:2005uw,2007JPhA...4013799C}. The Monte Carlo simulations for it were carried out in \cite{2007PhRvL..98c0602D}, indicating the second-order phase transitions in $d=3,4,5$. We will show that the estimates of critical exponents based on the two-sided Pad\' e extrapolations, using the $6-\epsilon$ and $2+\epsilon$ expansions, are in good agreement with the Monte Carlo results. 

The OSp$(1|2M)$ lattice systems may be constructed analogously: in this case there are $M$ pairs of Grassmann variables on each lattice site, $\theta_x^i, \bar\theta_x^i, i=1, \ldots M$, and
\begin{align}
\sigma_x = \left (1- 2\sum_{i=1}^M  \theta_x^i \bar \theta_x^i \right )^{1/2} \ , \qquad 
u_x\cdot u_y = - \sigma_x \sigma_y - \sum_{i=1}^M (\theta_x^i  \bar \theta_y^i +  \theta_y^i  \bar \theta_x^i) \ .
\end{align}
It is of obvious interest to study possible critical behavior in such lattice systems for $M>1$.
In this paper, we propose an extention of the field theoretic approach (\ref{cubicOSp}) to these systems. We are led to consider theories with interactions of order $2M+1$, i.e. $\sigma^{2M+1}$ plus terms involving the anticommuting fields.\footnote{Field theories with interactions involving all even powers of the field were studied long ago starting with \cite{Lipatov:1976ar}.}
Such theories have the upper critical dimensions
\be d_c(M) = 2 \frac{2M+1}{2M-1}\ .
\ee 

We use the results \cite{Gracey:2017okb} for the O$(N)$-invariant field theories with interactions of order $2M+1$, which are renormalizable at the upper critical dimensions $\frac{10}{3}$, $\frac{14}{5}$, etc., and substitute $N=-2M$ to account for the anticommuting nature of the $N$ scalar fields. Then we find OSp$(1|2M)$ invariant IR fixed points where the interaction term is proportional to $\left (\sigma^2+ 2\theta^i \bar \theta^i \right )^{(2M+1)/2}$ with an imaginary coefficient. These critical theories appear to be non-perturbatively well-defined, and it would be very interesting to compare the continuum results with those in the OSp$(1|2M)$ lattice systems.

\section{Scaling dimensions for the OSp$(1|2)$ model} 

The one-loop beta functions and anomalous dimensions for the theory (\ref{OSpfour}) are \cite{Fei:2015kta}
\begin{align}
\beta_1 &= -\frac{g_1 \epsilon}{2} -
 \frac{1}{12 (4\pi)^3} g_1 ( 10 g_1^2 + 12 g_1 g_2 - g_2^2 )\ , \notag \\
\beta_2 &= -\frac{g_2 \epsilon}{2}+ \frac{1}{4 (4\pi)^3}
(8 g_1^3 -2 g_1^2 g_2 -3 g_2^3 ) \ ,\notag \\  
\gamma_\theta &=   \frac{g_1^2}{6 (4\pi)^3} \ , \qquad 
\gamma_\sigma =   \frac{g_2^2 - 2 g_1^2}{12 (4\pi)^3}\ .
\end{align}
There is an OSp$(1|2)$ invariant IR fixed point where \footnote{There is another fixed point where the couplings have the opposite sign. The two are physically equivalent because they are related by the transformation $\sigma\rightarrow -\sigma$ in the path integral.}
\be
g_1= i \sqrt{(4\pi)^3\epsilon\over 5}\ , \qquad g_2= 2 g_1\ , \qquad \gamma_\theta= \gamma_\sigma =- \frac{\epsilon}{30}\ .
\ee
These results can be extended to the 4-loop order using the formulae from \cite{Gracey:2015tta} for the $O(N)$ invariant cubic model \cite{Fei:2014yja,Fei:2014xta}, and then setting $N=-2$. \footnote{The 5-loop renormalization of cubic theories in $6-\epsilon$ dimensions was carried out recently \cite{Borinsky:2021jdb,Kompaniets:2021hwg}, but we will not include 5-loop corrections here.} 
Then we find
\begin{equation}
\Delta_{\theta}= 2 - \frac{8}{15}\epsilon - \frac{7}{450} \epsilon^2-\frac{269 - 702 \zeta(3)}{33750} \epsilon^3-
\frac{207313 - 4212 \pi^4 + 936 \zeta(3) + 907200 \zeta(5)}{24300000} \eps^4  +\mathcal{O}(\eps^5)\ .
\label{fourloopdim}
\end{equation}
This expansion coincides with the corresponding one in the formal $q\rightarrow 0$ limit of $q$-state Potts model. 

The $2+\eps$ expansion for the $O(N)$ sigma model is \cite{Brezin:1975sq}
 \begin{equation}
\Delta_{\theta} = \frac{1}{2}\epsilon +\frac{1}{2(N-2)}\epsilon- \frac{N-1}{2(N-2)^2}\epsilon^2 +
\mathcal{O}(\eps^3)\ ,
\label{sigmadim}
\end{equation}
which for $N=-1$ becomes
\begin{equation}
\Delta_{\theta} = \frac{1}{3}\epsilon +\frac{1}{9}\epsilon^2 +
\mathcal{O}(\eps^3)\ .
\label{sigmadimm}
\end{equation}
Using the ``two-sided" $(4,3)$ and $(6,1)$ Pad\' e approximations, which utilize both the $6-\eps$ and $2+\eps$ expansions, we obtain the estimates
$\Delta_{\theta}\approx 1.46$ in $d=5$; $\Delta_{\theta}\approx 0.92$ in $d=4$
and $\Delta_{\theta}\approx 0.415$ in $d=3$. They are in very good agreement with the Monte Carlo simulations  of
the $q=0$ Potts model. Using the values for $\gamma/\nu$ given in Table I of \cite{2007PhRvL..98c0602D}, we find that 
\be
\Delta_\theta= \frac{d}{2}- \frac{\gamma}{2\nu} 
\ee
are very close to our Pad\' e estimates in $d=3,4,5$.

The next important operator in the sigma model has dimension \cite{Brezin:1975sq}
\begin{equation}
\Delta_{+} = d-\nu^{-1}= 2 - \frac{1}{N-2}\epsilon^2 +
\mathcal{O}(\eps^3)\ .
\label{nextdim}
\end{equation}
For $N=-1$ this becomes
\begin{equation}
\Delta_{+} = 2 + \frac{1}{3}\epsilon^2 +
\mathcal{O}(\eps^3)
\label{nextdimm}
\end{equation}
The expansion of $\Delta_+$ in $d=6-\epsilon$ may be found using the theory (\ref{cubicOSp}), where it corresponds to the OSp$(1|2)$ invariant primary operator 
$\sigma^2+ 2\theta \bar \theta$. Using the 4-loop results \cite{Gracey:2015tta}, we find
\be
\label{dplusdim}
\Delta_+= 4-\frac {2}{3}\eps
+ \frac{1}{30}\eps^2   +\frac{173 - 864 \zeta(3)}{27000} \epsilon^3+
\frac{51683 - 1296 \pi^4 + 140400 \zeta(3) + 272160 \zeta(5)}{4860000}\epsilon^4
  +\mathcal{O}(\eps^5)\ .
\ee
Performing the two-sided Pad\' e approximations, we find 
$\Delta_{+}\approx 3.36$ in $d=5$; $\Delta_{+}\approx 2.8$ in $d=4$
and $\Delta_{+}\approx 2.2$ in $d=3$. They are in good agreement with the Monte Carlo results for $\nu$ given in Table I of \cite{2007PhRvL..98c0602D} for $q=0$.

\section{Field theory for $M>1$}

The 2-d sigma model with target space H$^{0|2M}$ may be defined by picking one of the two solutions of the constraint
\be \sigma^2+ 2\sum_{i=1}^M \theta^i \bar \theta^i = 1\ .
\ee
The sigma model classical action is
\begin{align}
\label{sigmaOSpgen}
S=\frac {1} {2 g^2}\int d^2 x \bigg ( (\partial_\mu \sigma)^2 +
2\sum_{i=1}^M \partial_\mu \theta^i \partial^\mu \bar \theta^i
\bigg )\ ,
\end{align}
with the substitution of
\be 
\label{genconstr}
\sigma=\left ( 1- 2\sum_{i=1}^M \theta^i \bar \theta^i\right )^{1/2}
\ .
\ee
For example, for $M=2$
\be
\sigma=1-  \theta^1 \bar \theta^1 -\theta^2 \bar \theta^2  -  \theta^1 \bar \theta^1 \theta^2 \bar \theta^2\ .
\ee
In general, the expansion of the square root in (\ref{genconstr}) terminates with the term of order $2M$ proportional to $\prod_{i=1}^M \theta^i \bar \theta^i$. Related to this fact, we will propose a critical field theory
with interactions of order $2M+1$.

The H$^{0|2M}$ sigma model may be thought of as the O$(N)$ sigma model with $N=1-2M$. For $g^2<0$ it is asymptotically free in $d=2$ and therefore has a UV fixed point in $d=2+\epsilon$.
An interesting question is how to continue this theory to the dimension slightly below the upper critical one. For $M=1$ its upper critical dimension is $6$, and in $d<6$ we can view it as Euclidean field theory (\ref{cubicOSp}) with interaction $\left (\sigma^2+ 2\theta^i \bar \theta^i \right )^{3/2}$. However, such a description cannot be applicable to $M>1$. Indeed, already for $M=2$ the potential would contain the term $\sim   \theta^1 \bar \theta^1 \theta^2 \bar \theta^2/\sigma$ which is not admissible in renormalizable field theory. 
We propose that the proper generalization of our $M=1$ construction to higher $M$ involves higher powers in the OSp$(1|2M)$ invariant potential, so that its expansion in the anticommuting variables does not contain any terms with negative powers of $\sigma$; namely, 
\be
\left (\sigma^2+ 2\theta^i \bar \theta^i \right )^{(2M+1)/2}= \sigma^{2M+1}+ (2M+1) \sigma^{2M-1} \sum_{i=1}^M \theta^i \bar \theta^i + \ldots+ 
(2M+1)!!\ \sigma \prod_{i=1}^M \theta^i \bar \theta^i\ .
\ee
In $d_c(M)-\epsilon$ there is a weakly coupled IR fixed point of the interacting field theory
\be
\label{higherOSp}
S=\int d^d x \bigg (
\partial_\mu \theta^i \partial^\mu \bar \theta^i + \frac{1}{2}\left(\partial_{\mu}\sigma\right)^2 + g \left (\sigma^2+ 2\theta^i \bar \theta^i \right )^{(2M+1)/2} 
\bigg )
\ ,\ee
where at the fixed point $g$ is imaginary and $\sim \sqrt{\epsilon}$.

The $M=2$ model is particularly interesting, since its critical dimension $\frac{10}{3}$ is above $3$. In the $\frac{10}{3}-\epsilon$ expansion, we expect good results since $\epsilon=1/3$ is small. This was indeed the case for the model which is described by the $i\sigma^5$ theory. The $\frac{10}{3}-\epsilon$ expansion for this model was obtained in \cite{Codello:2017epp}, 
where it was called the Blume-Capel or the tricritical Lee-Yang universality class. \footnote{Quintic field theories with $S_q$ symmetry were similarly studied in \cite{Codello:2020mnt}.}

The quintic model with four anticommuting scalars and $Sp(4)$ symmetry has the Euclidean action
\be
\label{OSpfour}
S=\int d^d x \bigg (
\partial_\mu \theta^i \partial^\mu \bar \theta^i + 
\frac{1}{2}\left(\partial_{\mu}\sigma\right)^2 + \frac{g_1}{6}  \sigma \left ( \theta^i \bar \theta^i\right )^2
+ \frac{g_2}{6}  \sigma^3 \theta^i \bar \theta^i +  \frac{g_3}{120}  \sigma^5   
\bigg )
\ .\ee
Since
the one-loop $\beta$ and $\gamma$ functions were calculated in \cite{Gracey:2017okb} for the quintic model with $N$ additional scalar fields and O$(N)$ symmetry, we may set
$N=-4$ to obtain the corresponding results for the theory (\ref{OSpfour}).
After performing a suitable multiplicative redefinition of the couplings, $g_j \rightarrow  A g_j$, which simplifies the formulae, we find the one-loop beta functions for the theory (\ref{OSpfour}):
\begin{align}
\beta_1 = -\frac{3 g_1 \epsilon}{2} -
 \frac{1}{80} & ( 1864 g_1^3 + 10080 g_1^2 g_2 + 8664 g_1 g_2^2 + 
    10800 g_2^3 -480 g_1 g_2 g_3 + 1620 g_2^2 g_3 - 3 g_1 g_3^2)\ , \notag \\
\beta_2 = -\frac{3 g_2 \epsilon}{2}+ \frac{1}{80}
& (1120 g_1^3 + 2888 g_1^2 g_2 + 10800 g_1 g_2^2 - 15192 g_2^3 \notag \\  -
  & 80 g_1^2 g_3 + 1080 g_1 g_2 g_3 - 8640 g_2^2 g_3 - 1251 g_2 g_3^2)\ , \notag \\
\beta_3 = -\frac{3 g_3 \epsilon}{2}+ \frac{1}{16} &
(640 g_1^2 g_2 - 4320 g_1 g_2^2 + 23040 g_2^3 + 8 g_1^2 g_3 + 
  10008  g_2^2 g_3 - 1377 g_3^3)\ .
\end{align}
The anomalous dimensions are
\begin{align}
\gamma_\theta =  \frac{1}{20} (-2 g_1^2 + 3 g_2^2)\ , \qquad 
\gamma_\sigma =  \frac{1}{80} (8 g_1^2 - 72 g_2^2 + 3 g_3^2)\ .
\end{align}
We find an IR stable fixed points where
\be
(g_1, g_2, g_3)=  \frac{i}{3}  \sqrt{\epsilon\over 139} (3, 2, 8)
\ .
\ee
The two anomalous dimensions are equal at this fixed point, $\gamma_\theta= \gamma_\sigma= \frac{1}{4170} \epsilon$, suggesting that the Sp$(4)$ symmetry is enhanced to OSp$(1|4)$. This is indeed the case, since the interaction in (\ref{OSpfour}) combines at the fixed point into $\sim i \left (\sigma^2+ 2\theta^1 \bar \theta^1 + 2\theta^2 \bar \theta^2 \right )^{5/2}$.

The IR scaling dimension is 
\be 
\Delta_\theta= \Delta_\sigma= \frac{d-2}{2} + \gamma_\theta= \frac{2}{3}- \frac{1042}{2085} \epsilon + O(\epsilon^2)\ . 
\ee
Substituting $\epsilon=1/3$ gives the answer $\Delta_{1-loop} = \frac{3128}{6255}\approx 0.50008$. This is extremely close to the free dimension $1/2$ due to the smallness of the coefficient of $\epsilon$ in the anomalous dimension. We can try to improve on this estimate using the two-sided Pad\' e including the $2+\epsilon$ expansion for the $N=-3$ sigma model:
\begin{equation}
\Delta (2+\epsilon)= \frac{2}{5}\epsilon + \frac{2}{25}\epsilon^2 +
\mathcal{O}(\eps^3)\ .
\label{sigmadim}
\end{equation}
A Pad\' e approximant for the scaling dimension as a function of $d$,
\be 
\Delta_{P}(d) = {16 (d-2) (8360 - 2091 d)\over 476500 + d (-192316 + 18813 d)},
\ee
is consistent with the expansions near $10/3$ and $2$ dimensions. It gives $\Delta_P(3)\approx 0.485$ which is somewhat lower than $\Delta_{1-loop}$.
It would be interesting to calculate the $O(\eps^2)$ correction to $\Delta$ using the 2-loop diagrams in the quintic theory (\ref{OSpfour}). We expect this to reduce the uncertainty in estimating the scaling dimension in $d=3$. It would be also useful to study the critical exponents using the Functional Renormalization Group approach \cite{Dupuis:2020fhh} to theory (\ref{OSpfour}).     

Let us also discuss the theory with Sp$(6)$ symmetry and seventh-order interactions:
\be
\label{OSpsix}
S=\int d^d x \bigg (\sum_{i=1}^2
\partial_\mu \theta^i \partial^\mu \bar \theta^i + 
\frac{1}{2}\left(\partial_{\mu}\sigma\right)^2 + \frac{g_1}{90}  \sigma \left ( \theta^i \bar \theta^i\right )^3
+ \frac{g_2}{36}  \sigma^3 \left ( \theta^i \bar \theta^i\right )^2
+ \frac{g_3}{120}  \sigma^5 \theta^i \bar \theta^i +  \frac{g_4}{5040}  \sigma^7   
\bigg )
\ .\ee 
Using Gracey's results \cite{Gracey:2017okb} for the $O(N)$ symmetric theory with seventh-order interactions, and continuing them to $N=-6$, we 
find an IR fixed point in $\frac{14}{5}-\epsilon$ dimensions. In the normalization of couplings such that
\begin{align}
\gamma_\theta &= \frac{1}{3024}\left
(24 g_1^2 - 200 g_2^2 + 45 g_3^2\right)\ , \\
\gamma_\sigma &=\frac{1}{6048} \left (-48 g_1^2 + 1800 g_2^2 - 1350 g_3^2 + 15 g_4^2 \right )\ ,
\end{align}
it is located at
\be
(g_1, g_2, g_3, g_4)=i\cdot 0.00392514  \sqrt{\epsilon}  (15, 6, 8, 48)
\ .
\ee
At this fixed point
$\gamma_\theta=\gamma_\sigma\approx -5.50241\cdot 10^{-6}\epsilon$, and we
observe the enhancement of the Sp$(6)$ symmetry to OSp$(1|6)$. 

Our calculations provide evidence for the consistency of our proposal for $M>1$, but a lot remains to be done. In particular, 
it would be interesting to formulate a Monte Carlo approach to the OSp$(1|4)$ lattice model, where all the integrations in the partition function are over the Grassmann variables $\theta_x^1, \bar\theta_x^1, \theta_x^2, \bar\theta_x^2$. The results could be then compared with our prediction that the scaling exponents have exactly mean field values in four dimensions, but exhibit small deviations from them in three dimensions.  
Also, perhaps a conformal bootstrap approach to the critical exponents can be attempted along the lines of the method
\cite{Gliozzi:2013ysa}, which is applicable to non-unitary theories.

\section*{Acknowledgments}

I am very grateful to Roland Bauerschmidt and Tom Spencer for the important recent discussions and explanations that have rekindled my interest in field theories with OSp$(1|2M)$ symmetry, and for their comments on a draft of this paper. I am also grateful to Simone Giombi and Grisha Tarnopolsky for very useful discussions, and to them and Lin Fei for an earlier collaboration. I thank John Gracey for useful correspondence. Finally, I would like to thank Sergio Caracciolo and Giorgio Parisi for sharing their wisdom about the statistical mechanics of spanning forests.
This work was supported in part by the US NSF under Grant No.~PHY-1914860.

%%%%%%%%%%%%%

\bibliographystyle{ssg}
\bibliography{Critical}

\end{document}